\documentclass[a4paper]{article}

\usepackage[latin1]{inputenc}
\usepackage[english]{babel}
\usepackage[final]{graphicx}
\usepackage{verbatim}

\usepackage{amsmath,amsfonts,amssymb,amsthm}
\usepackage{latexsym}

\newcommand{\structure}[1]{$^{#1}$}

\newcommand{\authorstructure}[2][]{$ $\hspace{-5mm}$^{#1}$ {\small \it #2}}

\newenvironment{Aabstract}
	{\vspace{5mm}
	 \begin{center}\bf Abstract\end{center}
	 \begin{quote}\small}
	{\end{quote}\vspace{5mm}}

\begin{document}

\begin{center}
{\LARGE Gleam: the GLAST Large Area Telescope Simulation Framework}
\\[5mm]
{\sc Praveen Boinee\structure{a,b},
Giuseppe Cabras\structure{a,b},
Alessandro De Angelis\structure{a,b},
Dario Favretto\structure{a,c},
Marco Frailis\structure{a,b}, 
Riccardo Giannitrapani\structure{a,b},
Edoardo Milotti\structure{a,b},
Francesco Longo\structure{d},
Monica Brigida\structure{e},
Fabio Gargano\structure{e}, 
Nicola Giglietto\structure{e},
Francesco Loparco\structure{e},
Mario Nicola Mazziotta\structure{e},
Claudia Cecchi\structure{f},
Pasquale Lubrano\structure{f},
Monica Pepe\structure{f},
Luca Baldini\structure{g},
Johann Cohen-Tanugi\structure{g}, 
Michael Kuss\structure{g},
Luca Latronico\structure{g,h},
Nicola Omodei\structure{g,i},
Gloria Spandre\structure{g},
Joanne Bogart\structure{l},
Richard Dubois\structure{l},
Tune Kamae\structure{l},
Leon Rochester\structure{l},
Tracy Usher\structure{l},
Toby Burnett\structure{m},
Sean Robinson\structure{m},
Denis Bastieri\structure{n},
Riccardo Rando\structure{n}
}\end{center}

\authorstructure[a]{Dipartimento di Fisica, 
                   Universit\`a di Udine, 
                   via delle Scienze~208, 33100~Udine, Italy}

\authorstructure[b]{INFN,\:Sez.\:di Trieste,\:%
                   Gruppo Collegato di Udine,\:%
                   via delle Scienze\:208,\:33100~Udine, Italy}

\authorstructure[c]{CERN, CH-1211 Gen\`eve 23, Switzerland}

\authorstructure[d]{Dipartimento di Fisica, 
                   Universit\`a di Trieste \& INFN, Sez.~di Trieste, 
                   via Valerio~2, 34100~Trieste, Italy}

\authorstructure[e]{Dipartimento di Fisica, 
                   Universit\`a di Bari \& INFN, Sez.~di Bari,
                   via Orabona~4, 70126~Bari, Italy}

\authorstructure[f]{Dipartimento di Fisica, 
                   Universit\`a di Perugia \& INFN, Sez.~di Perugia,
                   via A.~Pa\-scoli, 06123~Perugia, Italy}

\authorstructure[g]{INFN, Sez.~di Pisa, 
                   via F.~Buonarroti~2, 56100~Pisa, Italy}

\authorstructure[h]{Dipartimento di Fisica, 
                   Universit\`a di Pisa,
                   via F.~Buonarroti~2, 56100~Pisa, Italy}

\authorstructure[i]{Dipartimento di Fisica, 
                   Universit\`a di Siena,  
                   via Roma~56, 53100~Siena, Italy}

\authorstructure[l]{Stanford Linear Accelerator Center,
                   2575 Sand Hill Road, MS 78,
                   Menlo Park, CA 94025, USA}

\authorstructure[m]{Physics Department,
                    University of Washington,
                    Box 351560, Seattle, WA 98195-1560,USA}

\authorstructure[n]{Dipartimento di Fisica, 
                   Universit\`a di Padova \& INFN, Sez.~di Padova,
                   via F.~Mar\-zolo~8, 35131~Padova, Italy}

\begin{Aabstract}
This paper presents the simulation of the GLAST high energy gamma-ray telescope. 
The simulation package, written in C++, is based on the Geant4 toolkit, and it is integrated into a general framework used to process events. A detailed simulation of the electronic signals inside Silicon detectors has been provided and it is used for the particle tracking, which is handled by a dedicated software. 
A unique repository for the geometrical description of the detector has been realized using the XML language and a C++ library to access this information has been designed and implemented. 
\end{Aabstract}

\section{Introduction}

The Gamma-ray Large Area Space Telescope (GLAST) is an international mission that will
study the high-energy phenomena in gamma-rays universe \cite{glast---,dubois}.
GLAST is scheduled for launch in 2006.
 
GLAST is instrumented with a hodoscope of Silicon planes with slabs of converter, followed by a calorimeter; the hodoscope is surrounded by an anticoincidence (ACD). This instrument, called the Large Area Telescope LAT, is sensitive to gamma rays in the
energy range between 30 MeV and 300 GeV.
The energy range, the field of view and the angular resolution of the GLAST LAT are vastly
improved in comparison with those of its predecessor EGRET (operating in 1991-2000), so that the LAT will provide a factor of 30 or more advance in sensitivity. 
This improvement should enable the detection of several thousands of new high-energy sources and allow the study of gamma-ray bursts and other transients, the resolution of the gamma-ray sky and diffuse emission, the search for evidence of dark matter and the detection of AGNs, pulsars and SNRs. One can find a detailed description of the scientific goals of GLAST mission and an introduction to the experiment in \cite{gsd}.

GLAST is a complex system, and detailed computer simulations are required 
to design the instrument, to construct the response function and to predict the background in the
orbit. To accomplish these tasks an object-oriented C++ framework called
{\it Gleam} (GLAST LAT Event Analysis Machine) was adopted and implemented. 

The main components Gleam are the subject of this paper.

\section{The GLAST Offline Software Framework}

\begin{figure}[!ht]
\centering
\includegraphics[angle=-90,width=0.85\textwidth]{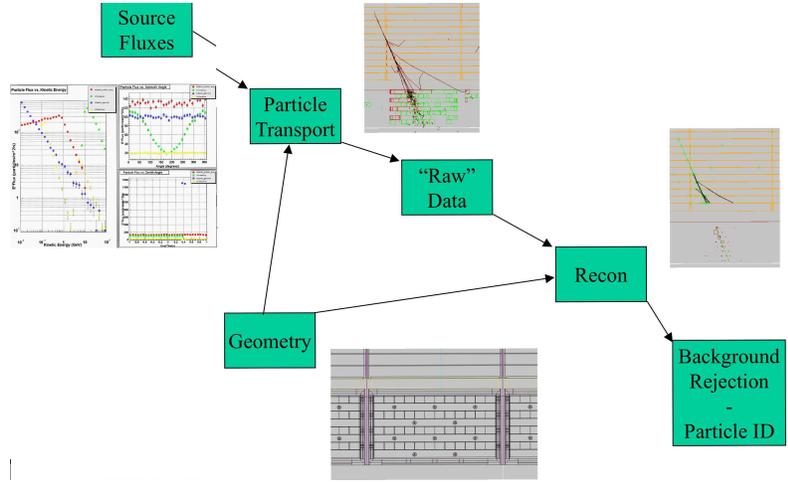}
\caption{General scheme for simulation and reconstruction within the GLAST off-line software framework}
\label{glast-sw}
\end{figure}

The structure of the GLAST offline software is described in figure~\ref{glast-sw}. 
Most packages has been developed explicitely by the GLAST collaboration for the specific items required by the simulation of a high energy gamma-ray telescope.
An important characteristics is the saparation of the packages according to their responsibilities.

The GLAST off-line software is based manly on Gaudi, a C++ framework, originally developed at CERN~\cite{Gaudi}.
Gaudi is an open project aimed to provide the necessary interfaces and services for building the framework of event data processing applications for High Energy Physics experiments.
A generic application framework should be usable for implementing the full range of offline computing tasks: generation of events, simulation of the detector, event reconstruction, testbeam data analysis, detector alignment, visualisation, etc. The framework gives the possibility to develop such tasks in a unified and coherent way.  
Tasks such as physics analysis and event reconstruction consist of the manipulation of mathematical or physical quantities (points, vectors, matrices, hits, momenta, etc.) by algorithms which are generally specified in terms of equations and natural language.  As a consequence, the Gaudi application framework makes a clear distinction between ``data'' objects and ``algorithm'' objects.

\begin{figure}[!hb]
\centering
\includegraphics[width=0.85\textwidth]{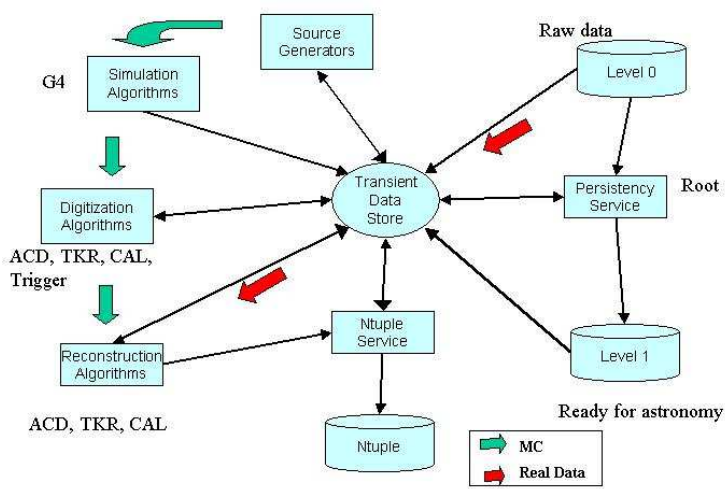}
\caption{Event processing within the GLAST off-line software framework}
\label{gaudi}
\end{figure}

In the GLAST framework, Gaudi manages the loop of particles to be simulated, then a series of algorithms are applied to each of them to get the result of the complete simulation and reconstruction chain. The figure~\ref{gaudi} shows the flux of simulated events within the Gleam framework based on Gaudi.

The output from the simulation is in the same format as the real data, and can be processed by the reconstruction package.

\section{The Source Generation package}

The Source Generation is the first algorithm called within the particle loop. Its task is to generate particles according to certain characteristcs. This algorithm must store the information on the temporal and spectral behaviour of the source, as well as on the orbital characteristics of GLAST. It provides a user interface to produce additional incoming particles and is responsible for setting the current time, the particle energy, direction, and type. Within this package a series of default sources are implemented. They include source for testing purposes as well as the description of astrophysical spectra and the expected particle and albedo gamma backgrounds.  

\begin{figure}[!hb]
\centering
\includegraphics[width=0.85\textwidth]{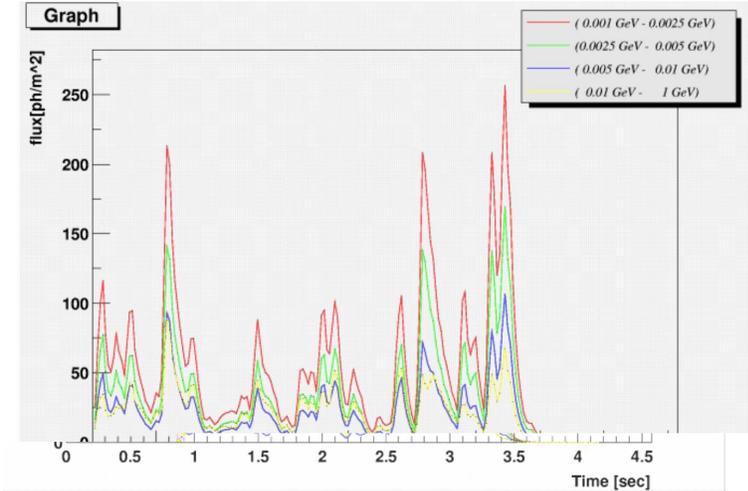}
\caption{An example of GRB light curve from the GLAST simulator.}
\label{fig--grb}
\end{figure}

An extension of this framework has been implemented for simulating transient sources such as Gamma-Ray Bursts (GRB). It can be used for studying the capability of GLAST for the observation of rapid transient fluxes in general. 
The physics adopted is based on the fireball model of Gamma Ray Burst, for which a series of shells is injected in the circumburst medium with different Lorentz factor~\cite{fireball}.
When a faster shell reaches a slower one a shock occours, and an accelerated electron distribution is obtained due to the shock acceleration process. Some of the energy dissipated during the shock is converted into a randomly oriented magnetic field. The electrons can loose their energy via synchrotron emission. The caratheristic synchrotron spectrum is boosted (and bemed) thancks to the Lorentz factor of the emittin matherial. The higher energy part of a GRB spectrum can be obtained keeping into account the possibility of Compton scattering of the synchrotron photons against the electron accelerated by the shock (Inverse Compton Scattering) \cite{grb--}. 
Figure~\ref{fig--grb} shows a light curve generated with this package.  

The information on the particles generated in the Source Generation package is put in the data store, so that every other piece of code in the framework can use it. 

\section{The Simulation package}

The algorithm which is responsible for generating the interactions of particles with the detector is based on the Geant4 MonteCarlo toolkit~\cite{geant4} which is an Object Oriented (OO) simulator of the passage of particles through matter. Its application areas include high energy physics and nuclear experiments, medical science, accelerator and space physics. 

Geant4 (G4) provides a complete set of tools for all the domains of detector simulation: Geometry, Tracking, Detector Response, Run, Event and Track management, Visualisation and User Interface. A large set of Physics Processes handle the diverse interactions of particles with matter across a wide energy range, as required by G4 multi-disciplinary nature; for many physics processes a choice of different models is available.
In addition a large set of utilities, including a powerful set of random number generators, physics units and constants, a management of particles compliant with the Particle Data Group, as well as interfaces to event generators and to object persistency solutions, complete the toolkit. G4 exploits advanced Software Engineering techniques and OO technology to achieve the transparency of the physics implementation and hence provide the possibility of validating the physics results. The OO design allows the user to understand, customise or extend the toolkit in all the domains. At the same time, the modular architecture of G4 allows the user to load and use only the components needed. To build a specific application the user-physicist chooses  among these options and implements code in user action classes supplied by the toolkit~\cite{geant4-intro}.

Within the Gleam framework the simulation is managed by the  Gaudi algorithm G4Generator~\cite{g4generator}. The main simulation is controlled by a customized and version of the G4 standard RunManager. Since the GLAST main event loop is driven by Gaudi and it will not use any graphics or data persistency features of Geant4, we have included in the RunManager only the real necessary parts for setup and run the generator. RunManager itself uses the following classes: 
\begin{itemize} 
\item DetectorConstruction: this class provides the list of materials and the geometry of the detector. In our case this information is stored in XML files; to access them the DetectorConstruction class uses methods of a Gaudi service. The Geometry class implements methods to traverse the geometry of GLAST and build a concrete Geant4 representation of it; the Material class does the same for the materials definitions.
\item PhysicsList: this class is the access point to the physics processes selection and customization. It uses other classes (GeneralPhysics, EMPhysics, HadronPhysics, MuonPhysics, IonPhysics) to set up particular physics processes. Since the Geant4 toolkit is open to new physics processes (along with new description of already present processes), this will be the access point for further development in the physics selection.
\item PrimaryGeneratorAction: this class is in charge of production and injection of primary particles in the detector simulation. In our cases it is linked to the Gaudi algorithm that is responsable to generate the incoming fluxes of particles.
\item DetectorManager: this class manages the setup and working of the sensitive detectors of the simulation and their interaction with the GAUDI Data Store. It is concretely implemented in the two subclasses PosDetectorManager and IntDetectorManager; the first one is associated with detector that saves hits information in the Si planes of the LAT tracker (TRK), while the second one is used for Anticoincidence (ACD) tiles and Calorimeter (CAL) cells.

\end{itemize}

\begin{figure}[!b]
\centering
\includegraphics[width=0.85\textwidth]{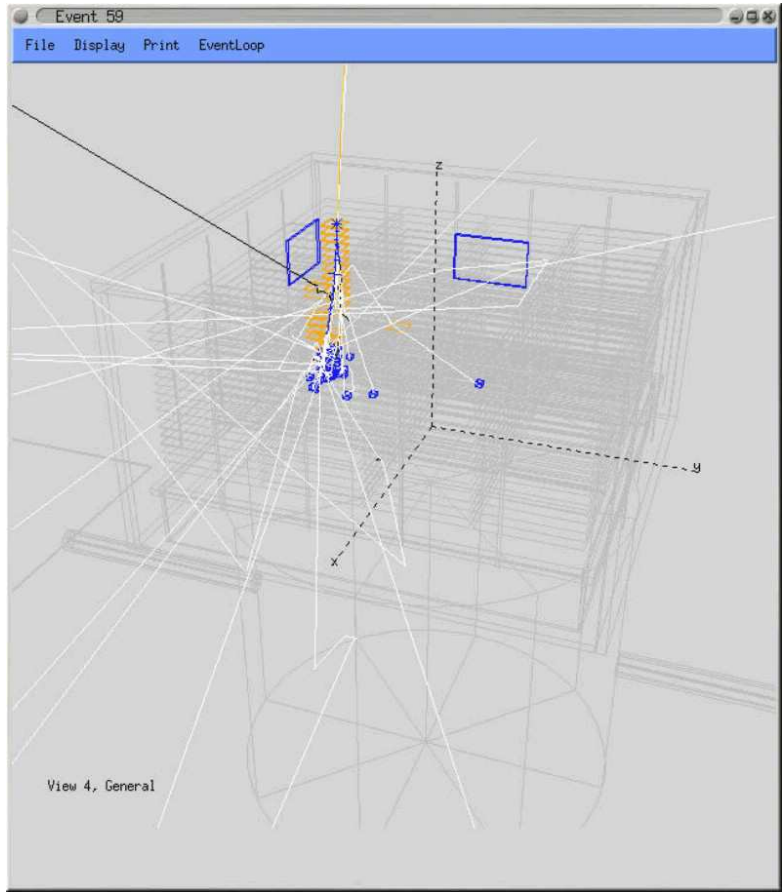}
\caption{10 MeV gamma interacting with the GLAST LAT detector}
\label{fig--gleam}
\end{figure}

Figure~\ref{fig--gleam} shows an event generated using Geant4 within the GLAST LAT experiment. 

\section{The Digitization package}

To implement a detailed digitization of the Tracker system a full simulation code has been developed. It takes into account all the main physical processes that take place in a
silicon strip detector (SSD) when it is crossed by an ionizing particle~\cite{bari}. 
The first version of the code has been written in FORTRAN and uses the HEED package for
simulating the energy loss of charged particles in silicon. The present version of the code has been
written in C++ and the process of energy loss is simulated by Geant4. 

\begin{figure}[!t]
\centering
\includegraphics[width=0.85\textwidth]{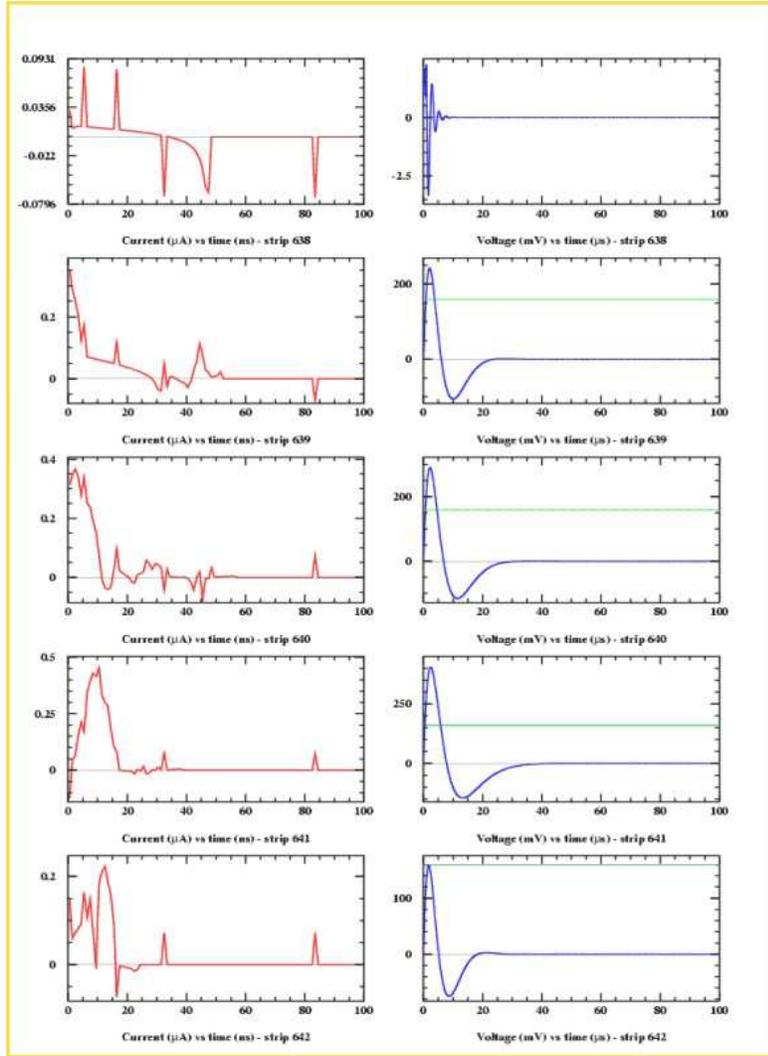}
\caption{Charge sharing for a 5 GeV electron, crossing the silicon 
 wafer at large zenith angle (60$^\circ$). The superimposed green  line
 represents the readout threshold voltage}
\label{fig--bari2}
\end{figure}

The input parameters of the code are the entry and exit points of the particle in a silicon ladder and the energy deposited by the particle, provided by the simulation package. Starting from these
parameters, the e-h pairs are generated along the track and are propagated towards the electrodes.
The current signals induced on each strip are evaluated and are
converted into voltage signals using the transfer function associated to the detector electronics, taking into account the
detector noise as well as the noise associated to the electronics. The
fired strips and the time over threshold (TOT) are then determined after imposing a threshold on the
voltage signals.
Figure~\ref{fig--bari2} shows some results of this package, concerning the signal generated in a silicon ladder equipped with the GLAST LAT electronics. 

Simplified version for the digitization of the signals are provided for the Calorimeter and the Anticoincidence panels. 

\section{The Reconstruction package}

This package contains the code that reconstructs tracks from hit strips in the LAT tracker. It's organized as a series of algorithms that act successively~\cite{tkrrecon}. 
Starting from digits generated by the Digitization package, it generates a series of clusters, that are used to find and fit the best track candidates.
This last procedure is done using alternative pattern recognition algorithms and a Kalman Filter based algorithm. Finally, using the best track found, another algorithm finds the best vertex candidate for gamma events.

\begin{figure}[!ht]
\centering
\includegraphics[width=0.85\textwidth]{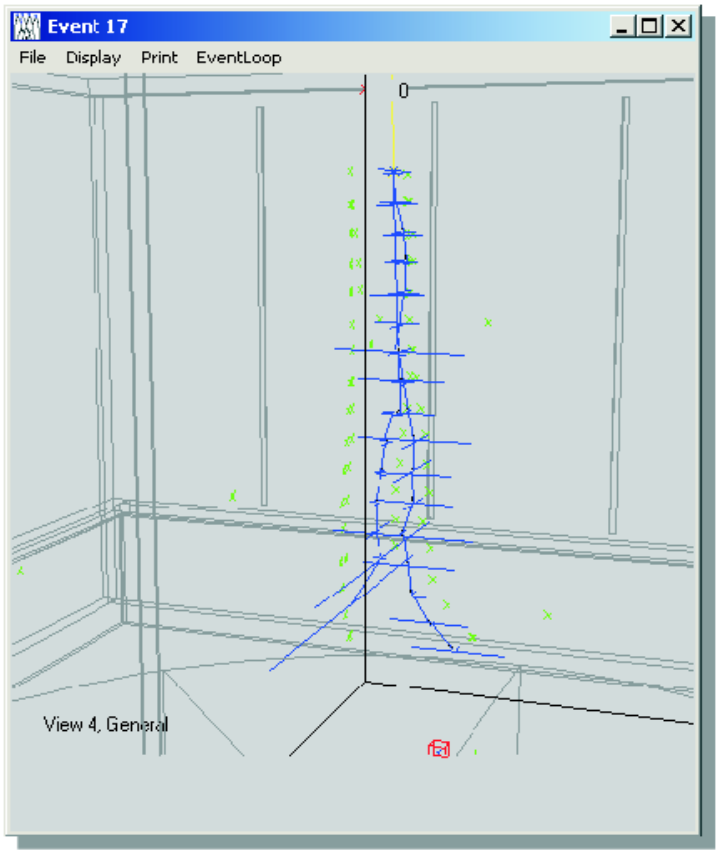}
\caption{100 MeV gamma tracked in the LAT}
\label{fig--tkrrecon2}
\end{figure}

Figure~\ref{fig--tkrrecon2} shows a photon track reconstructed in the GLAST LAT.

\section{Geometry database access library}

\begin{figure}
\centering
\includegraphics[width=0.85\textwidth]{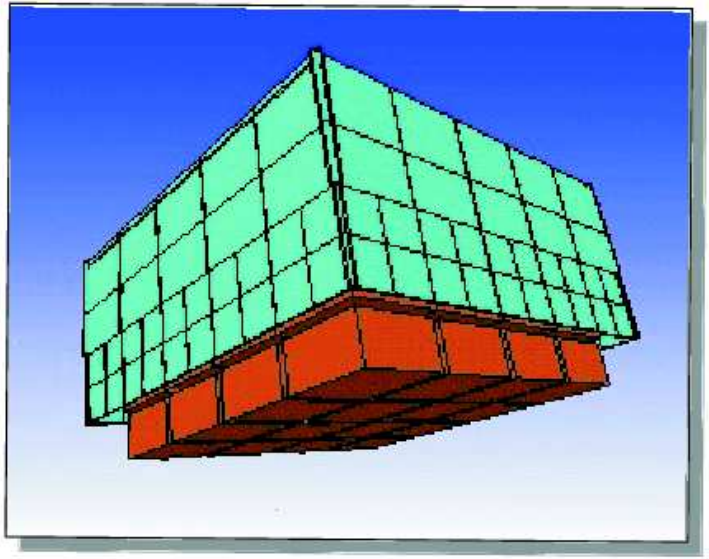}
\caption{GLAST LAT visualized with VRML}
\label{fig--vrml}
\end{figure}

The geometrical information of the detector and its materials are stored in an unique repository written using the XML language. A series of methods has been implemented to retrieve this information and make it available to the other packages. Interfaces have been implemented to visualize the geometrical information with different graphical outputs. Figure~\ref{fig--vrml} shows the external appearance of GLAST visualized with a VRML browser.  

\section{Event Display and GUI}

Although it is not part of the simulation, the visualization package is essential for the use of the simulation itself.

The Gleam event display GUI provides access to several kinds of controls. It includes controls on level of detail in detector display, rendering (or suppression) of digi products, recon products, choice of view, etc. It manages the Event generation with the possibility of choosing among many possible sources, ranging from simple test sources to simulation of albedo and galactic sources. 

A new version~\cite{Frailis:2003yp--,fred} of the event display based on the HepRep~\cite{Perl:2003wa--} protocol has been developed and will be soon integrated within the framework. 

\section{Conclusions}

The {\it Gleam} simulation program has been developed in the last few years and now it's ready for simulating the full GLAST satellite and will be used for deriving the final instrumental parameter and for generating a full set of events for the developing of scientific analysis software.


\begin{thebibliography}{99}

\bibitem{glast---}
{\tt http://glast.gsfc.nasa.gov/}

\bibitem{dubois} 
R.~Dubois, these proceedings

\bibitem{gsd} 
GLAST Science Brochure (March 2001),\\ {\tt http://glast.gsfc.nasa.gov/science/resources/}

\bibitem{Gaudi} 
Gaudi project, {\tt http://proj-gaudi.web.cern.ch/proj-gaudi/}

\bibitem{fireball} T. Piran (1999), Phys. Rep., {\bf 314}, 575

\bibitem{grb--} 
N.~Omodei and J.~Cohen-Tanugi,\\ {\tt http://www.pi.infn.it/$\sim$omodei/GRBSpectrum/}

\bibitem{geant4} 
S.~Agostinelli {\it et al.}, (2003),  NIM-A, {\bf 506}, 250

\bibitem{geant4-intro}
Geant4 toolkit home page: {\tt http://cern.ch/geant4}

\bibitem{g4generator} 
R.~Giannitrapani {\it et al.}, G4Generator, GLAST internal code review (2002)

\bibitem{bari} 
M.~Brigida {\it et al.} (2002), LAT internal note, LAT-TD-1058 

\bibitem{tkrrecon} 
B.~Atwood {\it et al.}, TkrRecon, GLAST internal code review (2002)

\bibitem{Frailis:2003yp--}
M.~Frailis and R.~Giannitrapani,
``The FRED Event Display: an Extensible HepRep Client for GLAST'', 
arXiv: {\bf cs.gr/0306031}

\bibitem{Perl:2003wa--}
J.~Perl, R.~Giannitrapani and M.~Frailis, ``The Use of HepRep in GLAST'', arXiv: {\bf cs.gr/0306059}

\bibitem{fred} 
M.~Frailis, R.~Giannitrapani, ``FRED: status of an Event Display for GLAST'', these proceedings 

\end{thebibliography}
\end{document}